\documentclass[12pt]{article}
\usepackage[dvips]{graphicx}
\usepackage{amssymb}
\usepackage{amsmath}

\def\beq{\begin{equation}}\def\eeq{\end{equation}}
\def\bea{\begin{eqnarray}}\def\eea{\end{eqnarray}}

\begin{document}

\title{Mensky's path integral and photon mass}
\author{Roman Sverdlov, Department of Mathematics, University of New Mexico} 

\date{August 10, 2016}
\maketitle

\begin{abstract}
\noindent It is commonly assummed that zero and non-zero photon mass would lead to qualitatively different physics. For example, massless photon has two polarization degrees of freedom, while massive photon at least three. This feature seems counter-intuitive. In this paper we will show that if we change propagator by setting $i \epsilon$ (needed to avoid poles) to a finite value, and also introduce it in a way that breaks Lawrentz symmetry, then we would obtain the continuous transition we desire once the speed of the photons is "large enough" with respect to "preferred" frame. The two polarization degrees of freedom will be long lived, while others will be short lived. Their lifetime will be near-zero if $m \ll \sqrt{\epsilon}$, which is where the properties of two circular polarizations arize.  The $i \epsilon$ corresponds to the intensity of Mensky's "continuous measurement" and the short lifetime of the longitudinal photons can be understood as the "conversion" of quantum degrees of freedom (photons) into "classical" ones by the measurement device (thus getting rid of the former). While the "classical" trajectory of the longitudinal photons does arize, it plays no physical role due to quantum Zeno effect: intuitively, it is similar to an electron being kept at a ground state due to continuous measurement. 
\end{abstract}

\subsection*{1. Introduction}

The idea of photon having mass has been around for some time (for a review of other people's thoughts on the topic, see \cite{PhotonMass}). This idea is quite appealing since it would lead to a lot of insides that are otherwise unavailable. For example, photon moves with speed near $c$ due to the randomly occuring forces action on it which, despite being very small, are a lot larger than the mass of the photon, leading to very large accelerations. Or, even if none of the forces were acting upon it, the near-lightlike velocity would be "most likely arbitrary close to $c$" in a sense that most of the randomly selected pairs of reference frames are moving with near-lightlike speed relative to each other. Since photon has such a small mass, it interacts with a lot of other "randomly occuring" sources \emph{after} it has been emitted and, therefore, its velocity is "independent" from the emitting source after short period of time, which makes it close to $c$. This new way of thinking of a photon also insipires a new version of Lawrentzian geometry that might be useful for causal set theory (see \cite{PhaseSpacetime}). 

However, one thing that stops us from picturing photon this way is the fact that massless photon only has two polarization degrees of freedom. On the other hand, if the photon had a mass, we would have been able to go to its reference frame, in which, due to spherical symmetry, we would have had \emph{at least} three degrees of freedom; the Lawrentz boost of the above would lead to three degrees of freedom in arbitrarily selected frame. So the question is: why would there be a fundamental difference between mass being \emph{very close to} zero and mass being \emph{exactly} zero? The answer is that if mass is exactly zero we won't be able to perform the above-mentioned Lawrentz boost from photons frame to our own, since its velocity will be exactly $c$. This, however, still feels a bit uncomfortable: intuitively we wish we could have physics that is continuous.

In order to get some inside as to where the problem is, let us rephrase what we have said about the boost. Let us assume that there \emph{is} some way of performing Lawrentz boost between lightlike and timelike frames. In this case we will arrive at a different problem: the boost in $z$-direction will lead to infinite stretch of $A^z$ (provided that photon moves in $z$- direction). But then how come we observe $A^z$ to be zero rather than infinitey? The answer is that photon "doesn't want" us to observe $A^z$ being infinite. Thus, it is "smart enough" to set $A^z=0$ in its own frame (it "knows" the observer will be moving relative to its frame in z- direction) which would lead to $A^z=0$ in our frame as well.  

However, there are two key assumptions that need to be noticed. From the point of view of reference frame of a photon, it sets $A^z$ to zero (as opposed to $A^x$ or $A^y$) because the observer moves in $-z$ direction (as opposed to $-x$ or $-y$). But what if there is a \emph{different} observer moving in $-x$ or $-y$ direction relative to the photon? Thus, we need to assume that \emph{such observers are not important}. This means that we set a "preferred" frame. This violation of relatity is "infinitesimal". After all, if Observer A moves in $-z$ direction relative to the photon, and Observer B moves with finite speed relative to Observer A, then Observer B also moves in $-z$ direction relative to the photon \emph{up to infinitesimal deviations in x- or y- directions}. Thus, \emph{if we neglect these infinitesimal deviations}, then the frames of Observer A and B would be "equally important" as far as photon is concerned. But, if we don't neglect them, then we will admit "infinitesimal" violation of relativity.

Now, even \emph{after} we admitted violation of relativity, we have yet another question to deal with. Namely, the value of $A^z$ in photons frame didn't have to be zero: it could have been infinitesimal! This would have allowed $A^z$ to have finite, yet non-zero, value in the frame of the "preferred" observer. The answer to this question is that, if the value of $A^z$ is restricted to "infinitesimal" range, we can't properly perform a path integral with respect to $A^z$. In other words, the $A^z$ would exist \emph{kinematically} but it would lack the dynamical properties and, therefore, we wouldn't be able to measure it. Our inability to perform path integral can be understood in terms of quantum Zeno effect. The photon is "continoususly measured" in such a way that $A^z$ is infinitesimal in photon's frame (or, equivalently, $A^z$ is finite in observer's frame) and this "continuous measurement" prevents $A^z$ from evolving. Now, since we have already admitted that observer's frame is a "preferred" one, we can claim that \emph{all} components of $A^{\mu}$ are "measured" in the observer's frame where the "measurement" is rather week: it keeps the finite but doesn't "glue" them to any specific finite value. Now, since $A^x$ and $A^y$ are \emph{naturally} finite, the measurement doesn't interfere with their natural evolution. On the other hand, since $A^z$ is naturally infinite, the measurement forces it to strongly deviate from its natural dynamics which is why we can't observe the latter. 

Now, our "trick" at allowing photon to be massive is to take the things we just admitted ("infinite" Laurentz stretch of $A^z$, "infinitesimal" violation of relativity and "infinitesimal" value of $A^z$) and replace the words "infinite", "infinitesimal" and "lightlike" with "large but finite", "small but finite", and "timelike, but very close to $c$", respectively. Thus, our new argument is that we have some large \emph{but finite} upper bound, $A_{max}$ on $\vert \vec{A} \vert$ which is imposed in \emph{preferred frame}. If we move relative to that preferred frame, the upper bound will no longer be spherically symmetric. However, this deviation would be small, provided that $A_{max}$ is large and velocities we are considering are not too close to light cone (this "smallness" of modification of $A_{max}$ corresponds to the "infinitesimal" violation of relativity previously discussed). Thus, in order to "respect" the $A_{max}$ in the observers frame, the photon has to limit the value of $A^z$ in its own frame to some small \emph{but finite} set of values -- the reason $A^z$ can be finite is that the speed of the photon is \emph{less than} $c$ (albeit very large).  

One way to justify the restriction on $\vert \vec{A} \vert$ is by referring to Mensky's idea of restricted path integral (\cite{Mensky1} \cite{Mensky2} \cite{Mensky3}). From Mensky's point of view, the probability amplitude of some "classical" trajectory, $A_{cl}^{\mu}$ taking place is the path integral over all trajectories $A^{\mu}$ that are lying "close enough" to $A_{cl}^{\mu}$:
\beq {\rm Amp} (A_{cl}) = \int [{\cal D} A] w (A, A_{cl})e^{iS(A)} \eeq
where $w (A, A_{cl})$ is some function that is near-unity when $A$ is close enough to $A_{cl}$ and near zero when $A$ is far away from $A_{cl}$.  In other words, the path integral is being taken around a "corridor" around $A_{cl}$. The size of the corridor  roughly corresponds to quantum-classical transition scale. Thus, on quantum scales the corridor \emph{appears} infinite, which explains why usually path integrals are taken without restrictions. On the other hand, on the classical scale the size of corridor appears to be so small that both of its walls "coincide" with one single "classical" trajectory, thus explaining classical physics.

Now, if we were to try to respect Lawrentz covariance, then the corridor would contain all $A^{\mu}$ satisfying $(A^{\mu} - A_{cl}^{\mu})(A_{\mu} - A_{cl ; \mu}) < \epsilon$. This would fill the vicinity of the lightcone of $A_{cl}$. Due to the non-compactness of the latter, the individual components of $A^{\mu}$ could be arbitrarily far away from $A_{cl}^{\mu}$. What makes the situation even worse is that the corridor is usually imposed by means of "weight function" $e^{-\frac{\epsilon}{2} \vert A - A_{cl} \vert^2}$. Now, if we are to respect relativity, then the argument under exponent would have both positive \emph{and} negative values. Thus, if the weight function would "suppress" the far away values of $A^0$, it would \emph{magnify} the far away values of $\vec{A}$ (and visa versa). To make the long story short, we have to violate relativity and impose a weight function of the form
\beq w (A, A_{cl}) = \exp \bigg( -\epsilon_1 (A^0 - A_{cl}^0)^2 - \epsilon_2 \sum_{k=1}^3 (A^k - A_{cl}^k)^2 \bigg) \label{NonRelativisticWeight} \eeq
with \emph{same} signs of both spacelike and timelike terms (the reason for $\epsilon_1 \neq \epsilon_2$ is simply that, since covariance is violated regardless, we don't see the reason for them to be equal; it will later turn out that $\epsilon_2 \gg \epsilon_1$ is "useful" choice in explaining lack of propagating longitudinal modes). Interestingly enough, the key ingredients of Mensky's path integral (imposition of $A_{max}$ and violation of relativity) are the same as the key ingredients that are needed for photon to have a mass! Thus, we will link the value of photon's mass to some power of the inverse of the size of Mensky's corridor. In particular, the former has to be "much smaller" than the latter in order for the third polarization degree of freedom (which is still present) to be negligible.

\subsection*{Mensky's integral and complex-valued mass} 

Before proceeding to spin 1 (in the sections that follow) let us illustrate our point on the example of spin 0 scalar Lagrangian, 
\beq {\cal L} = \frac{1}{2} \partial^{\mu} \phi \partial_{\mu} \phi - \frac{m^2}{2} \eeq
where the metric convention throughout the paper is assumed to be $(+ - - - )$. The effect of Mensky's modification is that the integrand is multiplied by an extra weight function of the form $\exp (- \epsilon (\phi - \phi_{cl})^2/2)$). Thus, the partition function is given by 
\beq Z (\phi_{cl}) = \int [{\cal D} \phi] w (\phi, \phi_{cl}) \exp \Bigg(i \int d^4 x \bigg(\frac{1}{2} \partial^{\mu} \phi \partial_{\mu} \phi - \frac{m^2}{2} \phi^2 \bigg) \Bigg) \eeq
where $w (\phi, \phi_{cl})$ is a \emph{weight function} given by 
\beq w (\phi, \phi_{cl}) = \exp \Bigg( -\frac{\epsilon}{2} \int d^4 x (\phi - \phi_{cl})^2 \Bigg) \label{WeightScalar} \eeq
This can be rewritten as 
\beq Z (\phi_{cl}) = \int [{\cal D} \phi] \exp \Bigg(i \int d^4 x {\cal L}\Bigg) \; , \; {\cal L} = \frac{1}{2} \partial^{\mu} \phi \partial_{\mu} \phi - \frac{m^2}{2} \phi^2 + \frac{i \epsilon}{2} (\phi - \phi_{cl})^2 \eeq 
The above expression for $\cal L$ can, after some simple algebra, be written as
\beq {\cal L} = \frac{1}{2} \partial^{\mu} \phi \partial_{\mu} \phi - \frac{m^2- i \epsilon}{2} \phi^2 - i \epsilon \phi_{cl} \phi + \frac{i \epsilon}{2} \phi_{cl}^2 \label{LagrangianFinal} \eeq
Now, this corresponds to the Lagrangian as we know it from "conventional" framework (such as \cite{Peskin}), with three modifications:

1) $m^2$ is replaced with $m^2 - i \epsilon$. 

2) The term $- i \epsilon \phi_{cl}$ corresponds to the "source" term $J \phi$.

3) There is an extra $\phi_{cl}^2$-term

If we absorb $i \epsilon$ into $m$, we would naturally have $i \epsilon$ in the denominator of propagator and avoid poles as desired. At the same time, $\epsilon$ is no longer infinitesimal since it is linked to the size of a corridor (or, in other words, the ratio between quantum and classical scales). Thus, we no longer "cheating just to avoid poles"; rather, $\epsilon$ has several different properties, albeit rather obscure ones. The $J$ corresponds to the "source term" in path integral interpretation of path integral. But in our case the interpretation of $J$ is different. Usually ppl think of $J$ as some set of particles or fields not included into the open quantum system that is being considered. In our case, we are \emph{only} talking about \emph{closed} quantum system that contains the enitre universe. Thus, $J$ is derived from the setup itself \emph{within} the framework of closed system (in particular, it is proportional to $\phi_{cl}$). This logically corresponds to the way we re-interpret the measurement: we don't view it as an interaction of open system with external sources but, rather, we view measurement (described by quantum corridors) as an \emph{internal} property of a \emph{closed} system. Finally, the $\phi_{cl}^2$ term might appear to produce unwanted deviation. But it had been shown in \cite{Epsilon} that said modification disappears if we remember to compute $\ln Z$ rather than $Z$. Another thing that should be noticed is that, as evident by $\partial^2 / \partial J (x_1) \partial J(x_2)$, the conventional QFT deals with infinitesimal deviations of $J$ from zero. In our case we view $J$ is finite. The correspondence with QFT is shown by the fact that the results for finite $J$ can be produced through "Taylor expansion" around $J=0$ where "derivatives" in Taylor expansion coincide with conventional QFT propagators. The fact that this is, in fact, the case has been shown in \cite{Epsilon}. 

However, it has also been noted in \cite{Epsilon} that the absorption of $i \epsilon$ into $m$ leads to an extra feature that does not occur in conventional physics: the particle has finite lifetime \emph{even if} it neither annihilates with, nor decays into, any other particles! Roughly speaking, this is due to the fact that $e^{i\omega t}$ is being replaced with $e^{it(\omega+i\epsilon/2\omega)} = e^{i\omega t} e^{- \epsilon t/2\omega}$; thus, the $e^{- \epsilon t/2 \omega}$ leads to attenuation. Now, if we consider the case of $t>0$, then we would close the contour of integration on the upper half of complex plane. In other words, we would select a pole $\omega + i \epsilon/2$, leading to the attenuation we just mentioned. On the other hand, if we take $t<0$, then we would close the contour of integration downwards, thus selecting the pole $-\omega-i \epsilon/2\omega$. This would lead to $e^{it(-\omega-i \epsilon/2\omega)}= e^{-i\omega t + t \epsilon/2\omega}$. Now, remembering that the above refers to $t<0$, we can combine it with $t>0$ result by writing $e^{i\omega \vert t \vert - \vert t \vert \epsilon/2\omega}$. This makes sense: after all, the order of derivatives shouldn't affect the final result.

The $e^{- \frac{\epsilon}{2} \vert t \vert^2}$ factor can be interpretted as a statement that if the particle exist "right now", then most likely it "did not" exist in the distant past, and "will not" exist in the distant future. In other words, it most likely had been "created out of nothing" and then later on it will "annihilate into nothing". Now, in conventional quantum field theory, the particles are being created by "sources" and annihilated by "sinks" (both corresponding to $J$-s). Said "sources" and "sinks" are interpretted by measurement events due to the interaction of open system with an environtment. In other words, conventionally, the particles are neither created, nor annihilated; rather, they are transformed from the environment into open system and then later return from open system back into environment. Now, in our model we replace open quantum system with the closed one. Thus, there is no such thing as "environment". Yet, we have measurement as internal property of closed system. This means that the ingredients of measurements should be replaced with their "equivalents" that would fit our viewpoint. The "transfering of particle from enviroment into the system" is replaced with "particle created out of nowhere" and "transfering of particle from system into environment" is replaced by "particle is annihilated into nowhere". Thus, the internal property of the particle has to be different: namely, it \emph{can} be created and destroyed without interaction with any other particles! It is simply that its lifetime will be very long (the smaller is $\epsilon$, the longer its lifetime would be) thus statistically it would be expected to be absorbed into some other particle before it "had a chance" to annihilate. 

In terms of Mensky's path integral, this can be understood in the following way. Since the weight function given in Eq \ref{WeightScalar} has integral sign, it would be $e^{- \infty} =0$ if $\phi - \phi_{cl}$ averages to some non-zero constant, however small it might be; this is due to the fact that the volume of spacetime is infinite. The only way for it to be non-zero is for $\phi - \phi_{cl}$ to approach zero assimptotically. At the same time it \emph{can} be far from zero throughout some region \emph{as long as} said region is limitted in spacetime. Roughly speaking, the deviation of $\phi - \phi_{cl}$ away from zero corresponds to some set of particles (which essentially amount to excitations of $\phi$), while the finiteness of the volume of that region implies that the lifetime of said particles is finite. One has to notice that $\phi_{cl}$ does \emph{not} approach $0$ at $\infty$, only $\phi - \phi_{cl}$ does. This means that $\phi_{cl}$ does \emph{not} consist of particles, only $\phi - \phi_{cl}$ does. In other words, particles constitute "quantum sector" while "classical sector" is particle free. The creation and annihilation of particles can be interpretted as interaction between quantum and classical sectors, which logically parallels the interaction between open system and environment in "conventional" interpretation. But the difference is that in conventional case both system and environment consist of particles, which makes it hard to draw the line between the two without violating mathematical rigour. In our case, quantum sector is the only part that consists of particles, which makes our theory far more acceptable for a mathematician. 

Physically, the attenuation is interpretted as the effect of "continuous measurement": since the particle corresponds to "quantum mechanical" degrees of freedom, it "disappears" once the system "collapses" into its classical state. The reason quantum field theory (which assumes non-attinuating particles) continues to be approximately valid is that the attinuation rate is much smaller than the rate at which the particle either decays or gets annihilated with some other particle. Now, even if it \emph{does} get absorbed into some other particle, it still won't escape the "ultimate annihilation" when either that new particle gets annihilate (or if that new particle also gets absorbed into some other particle then that "newer" particle might annihilate, and so forth). The necessity of "ultimate annihilation" is the result of the weight function we have just discussed. However, we can still claim that if we are computting a diagram consisting of small enough number of loops, this means that hte process under consideration would happen within a time interval much smaller than the time required for "ultimate annihilation". It is still possible that we are so "unlucky" that ultimate annihilation "interrupts" our diagram, however small it might be; but the probability of this is very small.

The key idea of this paper is that the statement about "very slow" rate of annihilation applies only to transverse photons but \emph{not} to longitudinal ones! After all, longitudinal photons, due to Lorentz transformation, would keep "hitting" our "corridor" which would cause them to be "measured" (and thus be subjected to possible annihilation) with far larger intensity. In other words, $\tau_{long} \ll \tau_{exp} \ll \tau_{transv}$ (where $\tau_{long}$ and $\tau_{transv}$ are lifetimes of non-interacting longitudinal and transverse photons, while $\tau_{exp}$ is a duration of typical experiment). The fact that $\tau_{exp} \ll \tau_{transv}$ implies that we can assume that non-interacting transverse photons have infinite lifetime, as we do in conventional calculations. On the other hand, the fact that $\tau_{long} \ll \tau_{exp}$ implies that typical longitudinal photon doesn't live long enough to interact with any other particle. \emph{That's why} it can never be detected! Now, from what we said earlier, the longitudinal photons don't disappear but, rather, they "get transferred into classical sector". But we have to remember that when we say "classical" we mean it in kinematic sense rather than dynamic. For example, when electron is kept in its ground state due to quantum Zeno effect, then its energy is "classical" in a sense that its uncertainty is near zero; but that doesn't change the fact that it doesn't participate in \emph{any} physical process -- \emph{either} classical \emph{or} quantum mechanical. Similarly, the longitudinal degrees of freedom, that might kinimatically exist, do not participate in physical processes either \emph{beyond} their \emph{very small} lifetime. 

We will now proceed to explicitly working out transverse and longitudinal photons in order to show that their lifetimes, in fact, end up being the way we wish them to be. 

\subsection*{Proposed modifications for electromagnetic Lagrangian}

As we have previously stated, $\epsilon$ is supposed to be included in the mass term of the Lagrangian. However, we have also pointed out in the Introduction that, due to the minus signs in Minskowskian metric, the "neighborhood" around any given $A_{cl}$ is not compact. Furthermore, a Laurentz covariant weight function, such as $w = \exp (\epsilon (A^{\mu} - A_{cl}^{\mu})(A_{\mu} - A_{cl; \mu}))$, can be both very large \emph{and} very small, depenting on whether expression under exponent happened to be spacelike or timelike. Thus, we need to break Lorentz covariance and use Eq \ref{NonRelativisticWeight} to define a weight function: 
\beq w (A, A_{cl}) = - \frac{\epsilon_1}{2} \int d^4 x (A^0 (x) - A_{cl}^0 (x) )^2 -  \frac{\epsilon_2}{2} \int d^4 x \vert \vec{A} (x)  - \vec{A}_{cl} (x) \vert^2 \eeq
Since the probability amplitude is computted by integrating $e^{iS}$, this would amount to the modifying Lagrangian by
\beq {\cal L}' = {\cal L} + \frac{i \epsilon_1}{2} (A^0 - A_{cl}^0)^2 + \frac{i \epsilon_2}{2} \vert \vec{A} - \vec{A}_{cl} \vert^2 \eeq
Now, similar to scalar case, we will attempt to absorb the above into mass term. Thus, we obtain
\beq  \frac{m^2}{2} g_{\mu \nu} A^{\mu} A^{\nu} + \cdots \longrightarrow \frac{m^2}{2} g_{\mu \nu} A^{\mu}A^{\nu} + \frac{i \epsilon_1}{2} (A^0)^2 + \frac{i \epsilon_2}{2} \vert \vec{A} \vert^2 + \cdots \label{MassTerm1} \eeq
Now, if we define $v^{\mu}$ to be a unit vector in time direction,
\beq v^{\mu} = \delta^{\mu}_0 \eeq
we then obtain 
\beq A^0 = v^{\mu} A_{\mu} \; , \; \vert \vec{A} \vert = \sqrt{(v^{\mu} A_{\mu})^2 - A^{\mu}A_{\mu}} \eeq
Thus, Eq \ref{MassTerm1} becomes
\beq {\cal L} = \frac{A^{\mu} A^{\nu}}{2} ((m^2 - i \epsilon_2) g_{\mu \nu} + i (\epsilon_1 + \epsilon_2) v_{\mu} v_{\nu} ) \eeq
If we now set 
\beq m_t^2 = m^2 + i \epsilon_1 \; , \; m_s^2 = m^2 - i \epsilon_2,  \label{mepsilon} \eeq
where $t$ and $s$ stand to "time" and "space", we obtain 
\beq {\cal L} = \frac{A^{\mu} A^{\nu}}{2} (m_s^2 g_{\mu \nu} + (m_t^2 - m_s^2) v_{\mu} v_{\nu}) + \cdots \eeq
Now, since we are taking a literalist stand when it comes to $\epsilon$, we will be consistent with our philosophy and also claim that the gauge fixing term, $(\partial^{\mu} A_{\mu})^2$, is to be taken literally as well. In the massless case, that term has no effect on equation of motion. In massive case, however, as we will see, it \emph{does} figure in the mass-related contributions. This is due to the fact that mass terms break $A^{\mu} \rightarrow A^{\mu} + \partial^{\mu} \Lambda$ symmetry. Therefore, we will include the gauge fixing term in the Lagrangian, and postulate
\beq {\cal L} = - \frac{1}{4} F^{\mu \nu} F_{\mu \nu} - \frac{1}{2 \xi} (\partial^{\mu} A_{\mu})^2 + ((m_t^2 - m_s^2) v_{\mu} v_{\nu} + m_s^2 g_{\mu \nu}) A^{\mu} A^{\nu} \eeq
as a complete electromagnetic Lagrangian. 

\subsection*{Complete set of photons}

Now we would like to find the Fourier transpose of the action. We have to be a little careful: in case of scalar field, we seem to be getting different signs depending on whether we use cosine or exponential:
\beq \phi = e^{ik^{\mu} x_{\mu}} \Longrightarrow \partial^{\mu} \phi \partial_{\mu} \phi = - k^{\mu} k_{\mu} \phi \eeq
\beq \phi = \cos (k^{\mu} x_{\mu}) \Longrightarrow \partial^{\mu} \phi \partial_{\mu} \phi = k^{\mu} k_{\mu} \phi \eeq
In order to resolve this, we note that in case of \emph{charged} scalar field, we have complex conjugation which results in sign being plus: 
\beq \phi = e^{ik^{\mu} x_{\mu}} \Longrightarrow \partial^{\mu} \phi^* \partial_{\mu} \phi = k^{\mu} k_{\mu} \phi \eeq
Thus, our "rule" is 
\beq \partial^{\mu} \partial^{\nu} \longrightarrow + k^{\mu} k^{\nu} \eeq
This produces an action
\beq S = \frac{A^{\mu} A^{\nu}}{2} \bigg((m_s^2 - k^2) g_{\mu \nu} + \bigg(1- \frac{1}{\xi}\bigg) k_{\mu} k_{\nu} + (m_t^2 -m_s^2) v_{\mu} v_{\nu} \bigg) \label{ActionToInvert} \eeq
Now we note that the Lagrange's equation is derived from variation of action being zero. This applies to \emph{all} variations that keep the boundary conditions fixed. In particular, this will be true if we change the \emph{magnitude} of the sinusoidal curve without changing the phase. The latter will produce
\beq \frac{\partial S}{\partial A^{\mu}} = (m_s^2 - k^2) A_{\mu} + \bigg(1- \frac{1}{\xi} \bigg) (A \cdot k) k_{\mu} + (m_t^2 - m_s^2) (A \cdot v) v_{\mu}  \label{LeastAction}\eeq
Equation this variation with zero implies that 
\beq A_{\mu} = \frac{1- \frac{1}{\xi}}{k^2 - m_s^2} (A \cdot k) k_{\mu} + \frac{m_t^2-m_s^2}{k^2-m_s^2} (A \cdot v) v_{\mu} \eeq
Now, since $\partial S/ \partial A^{\mu} = 0$, its contractions with $v^{\mu}$ and $k^{\mu}$ are likewise zero:
\beq 0 = v^{\mu} \frac{\partial S}{\partial A^{\mu}} = (m_s^2 - k^2) A \cdot v + \bigg(1- \frac{1}{\xi} \bigg) (A \cdot k) (k \cdot v)  + (m_t^2 - m_s^2) A \cdot v =  \label{vA} \eeq
\beq =  \bigg(1- \frac{1}{\xi} \bigg) (k \cdot v) (A \cdot k) + (m_t^2 - k^2) A \cdot v \eeq 
\beq 0 = k^{\mu} \frac{\partial S}{\partial A^{\mu}} = (m_s^2-k^2) A \cdot k + \bigg(1- \frac{1}{\xi} \bigg) (A \cdot k) k^2 + (m_t^2 - m_s^2) (A \cdot v) (k \cdot v) = \eeq
\beq = \bigg(m_s^2 - \frac{k^2}{\xi} \bigg) A \cdot k + (m_t^2 - m_s^2) (k \cdot v) (A \cdot v) \label{kA} \eeq
THe above two equations imply that 
\beq \left[ \begin{array}{cc}
\big(1- \frac{1}{\xi} \big) k \cdot v & m_t^2 - k^2 \\
m_s^2 - \frac{k^2}{\xi} & (m_t^2-m_s^2) k \cdot v \end{array} \right]
  \left[ \begin{array}{cc}
A \cdot k \\
A \cdot v \end{array} \right] =
\left[ \begin{array}{cc}
0 \\
0 \end{array} \right] 
 \eeq
If the above matrix has non-zero determinant, then the only solution is column vector being equal to zero:
\beq
 \left| \begin{array}{cc}
\big(1- \frac{1}{\xi} \big) k \cdot v & m_t^2 - k^2 \\
m_s^2 - \frac{k^2}{\xi} & (m_t^2-m_s^2) k \cdot v \end{array} \right| \neq  0
\Rightarrow 
 \left[ \begin{array}{cc}
A \cdot k \\
A \cdot v \end{array} \right] =
\left[ \begin{array}{cc}
0 \\
0 \end{array} \right] \eeq
The top component, $A \cdot k = 0$ implies that the wave is \emph{transverse}, while the bottom component, $A \cdot v =0$ implies that its direction does \emph{not} have timelike component in \emph{preferred time frame} defined by $v^{\mu} = \delta^{\mu}_{\nu}$. Furthermore, if we substitute the above into Eq \ref{LeastAction}, we obtain 
\beq 0 = \frac{\partial S}{\partial A^{\mu}} = (m_s^2 - k^2) A_{\mu} \Rightarrow \omega_{56} = \sqrt{m_s^2 + \vert \vec{k} \vert^2} \eeq
As we have stated, the non-zero determinant and, therefore, zero column vector, corresponds to \emph{transverse} photons. However, in case of massive photon, transverse waves are \emph{not} the only solution. We can produce \emph{longitudinal} photons if we set the determinant to zero which, ultimately, would allow column vector to be non-zero:
\beq
 \left[ \begin{array}{cc}
A \cdot k \\
A \cdot v \end{array} \right] \neq
\left[ \begin{array}{cc}
0 \\
0 \end{array} \right]
\Rightarrow
 \left| \begin{array}{cc}
\big(1- \frac{1}{\xi} \big) k \cdot v & m_t^2 - k^2 \\
m_s^2 - \frac{k^2}{\xi} & (m_t^2-m_s^2) k \cdot v \end{array} \right| = 0 \eeq 
If we now expand the above determant, we obtain 
\beq \bigg(1- \frac{1}{\xi} \bigg) (m_t^2-m_s^2) (k \cdot v)^2 - m_t^2m_s^2 + k^2 \bigg(m_s^2 - \frac{m_t^2}{\xi} \bigg) - \frac{k^4}{\xi} = 0 \label{det1} \eeq
We will now select a \emph{preferred} frame in which $t$-axis coincides with the direction of $v^{\mu}$:
\beq v^{\mu} = \delta^{\mu}_0 \Rightarrow k \cdot v = \omega \eeq
We will define $\vec{k}$ to be the projection of $k^{\mu}$ into $xyz$-plane in the above frame. Thus, 
\beq k^2 = \omega^2 - \vert \vec{k} \vert^2 \eeq
In this notation Eq \ref{det1} becomes 
\beq - \frac{\omega^4}{\xi} + \omega^2 \bigg( m_t^2 + \frac{m_s^2}{\xi} + \frac{2 \vert \vec{k} \vert^2}{\xi} \bigg) - m_t^2 m_s^2 - \vert \vec{k} \vert^2 \bigg(m_s^2 + \frac{m_t^2}{\xi} \bigg) - \frac{\vert \vec{k} \vert^4}{\xi} =0 \eeq
By solving it for $\omega^2$ and then taking square root, we obtain four solutions:  
\beq \omega_{1234} = \pm \sqrt{ \vert \vec{k} \vert^2 + \frac{m_s^2 + \xi m_t^2}{2} \pm \sqrt{ \vert \vec{k} \vert^2 (m_t^2 - m_s^2) (\xi -1) + \frac{1}{4} (\xi m_t^2 - m_s^2)^2}} \label{w1234} \eeq
Now, the solutions $\omega_{56}$ were corresponding to transverse wave. This is \emph{not} the case for $\omega_{1234}$. So let us find out the type of wave that $\omega_{1234}$ corresponds to. If we rewrite Eq \ref{vA} in terms of our "preferred" frame, we obtain
\beq \omega \bigg(1- \frac{1}{\xi} \bigg) (A^0 \omega - \vec{A} \cdot \vec{k}) + (m_t^2 + \vert \vec{k} \vert^2 - \omega^2) A^0 =0 \eeq
which, after simple algebra, becomes
\beq A^0 \bigg(m_t^2 + \vert \vec{k} \vert^2 - \frac{\omega^2}{\xi} \bigg) - \omega \bigg(1 - \frac{1}{\xi} \bigg) \vec{A} \cdot \vec{k} = 0 \eeq
which we then solve for $A^0$ and obtain 
\beq A^0 = \frac{\omega \big(1 - \frac{1}{\xi} \big) \vec{A} \cdot \vec{k}}{m_t^2 + \vert \vec{k} \vert^2 - \frac{\omega^2}{\xi}}  \eeq

\subsection*{Choice of $\epsilon$-s, $m$ and $\xi$ that would "hide" longitudinal modes}

We now recall that $m_s$ and $m_t$ are given by Eq \ref{mepsilon}. By substitutting this into Eq \ref{w1234} we obtain 
\beq \omega_{1234}^2 = \vert \vec{k} \vert^2 + m^2 + \frac{(\xi -1)m^2 + i (\xi \epsilon_1 - \epsilon_2)}{2} \pm \nonumber \eeq
\beq \pm \sqrt{i (\epsilon_1 + \epsilon_2)(\xi - 1) \vert \vec{k} \vert^2 + \frac{(m^2 (\xi -1) + i (\xi \epsilon_1 - \epsilon_2))^2}{4}} \label{AttenuationOmega} \eeq
In conventional QFT it is assumed that both $\epsilon$ and $m$ are zero. In our case we are saying that they are non-zero, but very small. Since they are dimensionful, the way to say that they are very small is to say that $\epsilon \ll \vert \vec{k} \vert^2$ and $m \ll \vert \vec{k} \vert$. Now, we "know" the range of values of $\vec{k}$. But we \emph{don't know} the way in which $\epsilon$ and $m^2$ compare to each other. Thus, we are to consider separately the case where $\epsilon \ll m^2$ and $m^2 \ll \epsilon$. Furthermore, we don't know the value of $\xi$ \emph{at all}. So we are to consider $\xi \ll 1$, $\xi \approx 1$ and $\xi \gg 1$. Also, if, for example, we want to say that $\xi \gg 1$ and $m^2 \gg \epsilon$, we are to deside whether $m^2/\epsilon \gg \xi$ or $\xi \gg m^2/ \epsilon$. And we are to ask similar question with regards to $(\xi -1)^{-1}$ and $\xi^{-1}$ when we make a choice of $\xi \approx 1$ and $\xi \ll 1$, respectively. And finally we have to also repeat all that for $m \ll \epsilon$ and $m \approx \epsilon$, as well. 

If we are to go through combinations of choices blindly, the list of combinations to consider would be very long. Therefore, we will shorten the list of combinations by keeping in mind what kind of $\omega_{1234}$ we are "trying" to get. In particular, our "goal" is to say that the reason we don't see longitudinal photons is simply that their lifetime is very short. In other words, if we could find a way to make $\omega_{1234}$ large, we would be able to say that longitudinal photons are either "too heavy to propagate" (if the real part of $\omega_{1234}$ is large) or they attenuate according to our "new" concept (if the imaginary part of $\omega_{1234}$ is large). In other words, as long as an \emph{absolute value} of $\omega_{1234}$ is large, we have one, or both, explanations as to why we wouldn't see longitudinal photons. This would ultimately explain why the number of \emph{long lived} degrees of freedom is $2$, \emph{even if} photon mass is non-zero, as long as said \emph{non-zero} photon mass compares to $\epsilon$ as well as some function of $\xi$ in the prescribed way we are about to discover. 

Now, it \emph{is} possible to detect a single wave length. This means that the lifetime of longitudinal photons has to be much smaller than the period of the wave. In other words, we need to have 
\beq \vert \omega_{1234} \vert \gg \vert \vec{k} \vert \label{goal} \eeq
Now, Eq \ref{AttenuationOmega} is written in such a way that the only thing that occurs in denominator are factors of $2$ and $4$. This means that none of the "small" parameters can possibly make $\omega_{1234}$ "large". Now, the "large" parameter $\vert \vec{k} \vert$ can't possibly make $\omega_{1234}$ larger than itself. Thus, we need some "large" parameter \emph{other than} $\vec{k}$ in order for Eq \ref{goal} to hold. Now, we already know that neither $m$ nor $\epsilon$ are "large" (in fact, they are very small). So the only candidate for "large" parameter is $\xi$. Thus, we conclude that 
\beq \xi \gg 1 \label{xi} \eeq
Now we can rewrite Eq \ref{AttenuationOmega} as 
\beq \omega_{1234}^2 = \vert \vec{k} \vert^2 + m^2 + \sqrt{P} \pm \sqrt{P+Q} \eeq
where
\beq P = \bigg( \frac{(\xi -1)m^2 + i (\xi \epsilon_1 - \epsilon_2)}{2} \bigg)^2 \; , \; Q = i (\epsilon_1 + \epsilon_2) (\xi -1) \vert \vec{k} \vert^2 \label{PQ} \eeq
Now, in the situation when we have $-$ in the place of $\pm$, we might have the kind of cancelation that we don't have in case of $+$ sign. Thus, it is conceivable that the two harmonics corresponding to + sign attenuate fast, while the ones corresponding to - sing attenuate slow. Now since \emph{all four} of them are longitudinal, we want \emph{all} of them to attenuate fast. Thus, we have to take care of the situation with - sing; and, if we succeed, the + case would work by default. Now, it is clear that the situation with minus sign produces $\omega_{1234} \gg \vert \vec{k} \vert^2$ if and only if
\beq \vert \sqrt{P + Q} - \sqrt{P} \vert \gg \vert \vec{k} \vert^2 \label{Puzzle} \eeq
Now let us denote the above difference by $R$:
\beq R = \sqrt{P+Q} - \sqrt{P} \eeq
This means that 
\beq P + Q = (R + \sqrt{P})^2 = R^2 + P + 2 R \sqrt{P} \eeq
and, therefore,
\beq Q = R^2 + 2R \sqrt{P} \eeq
By triangle inequality, this implies
\beq \vert Q \vert \leq \vert R \vert^2 + 2 \vert R \vert \sqrt{ \vert P \vert} \label{Triangle} \eeq
This implies that \emph{at least one} of the two terms on right hand side of Eq \ref{Triangle} has to be greater or equal to $\vert Q \vert/2$: 
\beq \bigg( \vert R \vert^2 \geq \frac{\vert Q \vert}{2} \bigg) \vee \bigg(2 \vert R \vert \sqrt{\vert P \vert} \geq \frac{\vert Q \vert}{2} \bigg) \eeq
which can be rewritten as 
\beq \bigg( \vert R \vert \geq \sqrt{\frac{\vert Q \vert}{2}} \bigg) \vee \bigg ( \vert R \vert \geq \frac{\vert Q \vert}{4 \sqrt{ \vert P \vert}} \bigg) \eeq
or, in other words,
\beq \vert R \vert \geq \min \bigg(\sqrt{\frac{\vert Q \vert}{2}}, \frac{ \vert Q \vert}{4 \sqrt{ \vert P \vert}} \bigg) \label{MinimumPrecurson} \eeq
Now, the \emph{sufficienty} condition for $\vert R \vert \gg \vert \vec{k} \vert$ is 
\beq \bigg[\min \bigg(\sqrt{\frac{\vert Q \vert}{2}}, \frac{ \vert Q \vert}{4 \sqrt{ \vert P \vert}} \bigg) \gg \vert \vec{k} \vert^2 \bigg] \Longrightarrow [\vert R \vert \gg \vert \vec{k} \vert^2] \label{Minimum} \eeq
Let us therefore try various ways of violating the assumption on the left hand side of Eq \ref{Minimum}. One way of doing that is to consider the negation of the "largeness" of $\sqrt{\vert Q \vert}$:
\beq \neg \bigg(\sqrt{\frac{\vert Q \vert}{2}} \gg \vert \vec{k} \vert^2 \bigg) \label{Negation1}\eeq
Now, from Eq \ref{Puzzle} together with triangle inequality we have 
\beq \vert R \vert \leq \sqrt{\vert P + Q \vert} + \sqrt{\vert P \vert} \leq \sqrt{\vert P \vert + \vert Q \vert} + \sqrt{\vert P \vert} \eeq
This implies that 
\beq R \gg \vert \vec{k} \vert^2 \Rightarrow [(\vert P \vert \gg \vert \vec{k} \vert^4) \vee (\vert Q \vert \gg \vert \vec{k} \vert^4)] \label{Or} \eeq
The combination of Eq \ref{Negation1} and \ref{Or} implies that 
\beq \vert P \vert \gg \vert \vec{k} \vert^4 \label{PisLarge} \eeq
The combination of Eq \ref{Negation1} and Eq \ref{PisLarge} implies 
\beq \vert P \vert \gg \vert Q \vert \label{Pdominates} \eeq
This, in turn, implies that 
\beq R = \frac{Q}{2 \sqrt{P}} \eeq
The Eq \label{Pdominates} them implies that 
\beq \vert R \vert  \ll \frac{\vert Q \vert}{2 \sqrt{\vert Q \vert}} = \frac{\sqrt{\vert Q \vert}}{2} \eeq
Which, per Eq \ref{Negation1} implies 
\beq \neg (\vert R \vert \gg \vert \vec{k} \vert^2) \label{Bad1} \eeq
Since our goal is Eq \ref{goal}, the above negation is precisely what we don't want! Therefore, the original assumption that we have made that lead us to Eq \ref{Bad1} has to be false. That assumption is Eq \ref{Negation1}. The falsity of that equation implies the truthfulness of the statement 
\beq \sqrt{\frac{\vert Q \vert}{2}} \gg \vert \vec{k} \vert^2 \label{Affirmation1}\eeq
Therefore, the only other way for the left hand side of Eq \ref{Minimum} to be false is to postulate a different negation:
\beq \neg \bigg( \frac{\vert Q \vert}{4 \sqrt{\vert P \vert}} \gg \vert \vec{k} \vert^2 \bigg) \label{Negation2}\eeq
This, together with Eq \ref{Affirmation1} implies that
\beq \sqrt{\frac{\vert Q \vert}{2}} \gg \frac{\vert Q \vert}{4 \sqrt{ \vert P \vert}} \eeq
By dividing both sides by $\vert Q \vert$ this becomes
\beq \frac{1}{\sqrt{2 \vert Q \vert}} \gg \frac{1}{4 \sqrt{\vert P \vert}} \eeq
which implies that 
\beq \vert Q \vert \ll \vert P \vert  \eeq
Once agian, this implies that 
\beq R = \frac{Q}{2 \sqrt{P}} \eeq
which, per Eq \ref{Negation2} implies that 
\beq \neg (R \gg \vert \vec{k} \vert^2) \eeq
which is again something we don't want. Thus, we have shown that, no matter how we try to make left hand side of Eq \ref{Minimum} false, we would always arrive at the negation of $R \gg \vert \vec{k} \vert^2$. Therefore, since we want $R \gg \vert \vec{k} \vert^2$ to be true, we are forced to assume that the left hand side of Eq \ref{Minimum} is true as well. This means that Eq \ref{Minimum} is \emph{necessary}; and its sufficiency we have already proven earlier. Thus left hand side of Eq \ref{Minimum} is both necessary and sufficient for $R \gg \vert \vec{k} \vert^2$ to hold.

Let us now go back to trying to find the relationship between $m$, $\epsilon_1$, $\epsilon_2$ and $\xi$. If we substitute Eq \ref{PQ} into left hand side of Eq \ref{Minimum}, we obtain
\beq \bigg(\sqrt{\frac{\vert \vec{k} \vert^2 (\epsilon_1 + \epsilon_2) (\xi -1)}{2}} \gg \vert \vec{k} \vert^2 \bigg) \wedge \bigg( \frac{\vert \vec{k} \vert^2 (\epsilon_1 + \epsilon_2)(\xi-1)}{\sqrt{m^4(\xi-1)^2 + (\xi \epsilon_1 - \epsilon_2)^2} } \gg \vert \vec{k} \vert^2 \bigg) \eeq 
Now we recall from Eq \ref{xi} that $\xi \gg 1$. Thus, we freely replace $\xi -1$ by $\xi$ and obtain 
\beq \bigg(\sqrt{\frac{\xi \vert \vec{k} \vert^2 (\epsilon_1 + \epsilon_2)}{2}} \gg \vert \vec{k} \vert^2 \bigg) \wedge \bigg( \frac{\vert \vec{k} \vert^2 (\epsilon_1 + \epsilon_2) \xi}{\sqrt{m^4 \xi^2 + (\xi \epsilon_1 - \epsilon_2)^2}} \gg \vert \vec{k} \vert^2 \bigg) \eeq
The reason we didn't replace $\xi \epsilon_1 - \epsilon_2$ is that it is conceivable that $\epsilon_2/ \epsilon_1 \gg 1$ and, in fact, said ratio could be comparable to $\xi$. As a matter of fact, we will soon show that this is indeed what we will be lead to have. Now, the "strong" inequalities we are dealing with are not affected by factor of $2$, so we can throw it away. Furthermore, we can cancel factors of $\vert \vec{k} \vert$ whenever possible and we can also do some squaring in order to get rid of square root signs. This leads to 
\beq (\xi (\epsilon_1 + \epsilon_2) \gg \vert \vec{k} \vert^2) \wedge ((\epsilon_1 + \epsilon_2)^2 \xi^2 \gg m^4 \xi^2 + (\xi \epsilon_1 - \epsilon_2)^2) \label{NoRoots} \eeq
It is easy to see that 
\beq \forall \alpha >0 \forall \beta >0 [(\gamma \gg \alpha + \beta) \Leftrightarrow ((\gamma \gg \alpha) \wedge (\gamma \gg \beta))] \eeq
The above statement is true only for positive $\alpha$ and $\beta$ because if it turns out that they have opposite signs it is possible that the magnitude of each of $\alpha$ and $\beta$ is large, while their difference is small (for example, one can have $\alpha = 10^6 + 10^{-6}$ and $\beta = 10^{-6} - 10^6$). But for positive $\alpha$ and $\beta$ the above statement is correct. Now, noticing that all of the ignredients of Eq \ref{NoRoots} are positive, we can rewrite it as 
\beq [\xi (\epsilon_1 + \epsilon_2) \gg \vert \vec{k} \vert^2 ] \wedge [(\epsilon_1 + \epsilon_2)^2 \xi^2 \gg m^2 \xi^2] \wedge [(\epsilon_1 + \epsilon_2)^2 \xi^2 \gg (\xi \epsilon_1 - \epsilon_2)^2] \eeq
which can be simplified as 
\beq [\xi (\epsilon_1 + \epsilon_2) \gg \vert \vec{k} \vert^2 ] \wedge [\epsilon_1 + \epsilon_2  \gg m^2 ] \wedge [(\epsilon_1 + \epsilon_2) \xi \gg \xi \epsilon_1 - \epsilon_2] \label{Triade} \eeq
Let us evaluate the right hand side. First, we will divide it by $\xi$, thus obtaining
\beq \epsilon_1 + \epsilon_2 \gg \epsilon_1 - \frac{\epsilon_2}{\xi} \eeq
Now, it is easy to see that 
\beq \bigg[ \epsilon_1 + \epsilon_2 \gg \epsilon_1 - \frac{\epsilon_2}{\xi} \bigg] \Leftrightarrow \bigg[ (\epsilon_1 + \epsilon_2 \gg \epsilon_1) \vee \bigg(\epsilon_1 \gg \epsilon_1 - \frac{\epsilon_2}{\xi} \bigg) \bigg] \eeq 
Now one can see that 
\beq \epsilon_1 \gg \epsilon_1 - \frac{\epsilon_2}{\xi} \Leftrightarrow \frac{\epsilon_2}{\xi} \approx \epsilon_1 \Rightarrow \epsilon_2 \gg \epsilon_1 \eeq
And also 
\beq \epsilon_1 + \epsilon_2 \gg \epsilon_1 \Leftrightarrow \epsilon_2 \gg \epsilon_1 \eeq
In other words, we have the following logical structure 
\beq Z \Leftrightarrow X \vee Y \; , \; Y \Rightarrow X \eeq
It is easy to see that this implies
\beq Z \Leftrightarrow X \eeq
In other words, 
\beq \bigg(\epsilon_1 +\epsilon_2 \gg \epsilon_1 - \frac{\epsilon_2}{\xi} \bigg) \Leftrightarrow (\epsilon_2 \gg \epsilon_1) \eeq
Thus, we can rewrite Eq \ref{Triade} as 
\beq (\xi (\epsilon_1 + \epsilon_2) \gg \vert \vec{k} \vert^2 ) \wedge (\epsilon_1 + \epsilon_2 \gg m^2) \wedge (\epsilon_2 \gg \epsilon_1) \eeq
We can now use $\epsilon_2 \gg \epsilon_1$ to simplify the statements in the left and in the middle, to obtain 
\beq (\xi \epsilon_2 \gg \vert \vec{k} \vert^2) \wedge (\epsilon_2 \gg m^2) \wedge (\epsilon_2 \gg \epsilon_1) \eeq
Now, the statement at the left hand side can't possibly be true for \emph{all} $\vec{k}$, since $\vert \vec{k} \vert$ can be arbitrarily large, while the constants $\xi$ and $\epsilon_2$ are fixed. To solve this problem we again appeal to our literalist view and claim that \emph{ultraviolet cutoff } $\Lambda$ that is normally assumed to be a \emph{flowing} large number is actually \emph{fixed}, and we will use the value of $\Lambda$ as an upper bound on $\vert \vec{k} \vert$. Thus, we say
\beq (\xi \epsilon_2 \gg \Lambda^2) \wedge (\epsilon_2 \gg m^2) \wedge (\epsilon_2 \gg \epsilon_1) \eeq
One has to be a little careful here since, strictly speaking, $\Lambda$ is \emph{not} the upper bound on momentum \emph{itself} but rather on a \emph{change} of momentum at a junction of Feynmann diagram; thus if a diagram has few loops it is possible for momentum to exceed $\Lambda$. However, since we put $\gg$ sing instead of simple $>$, the statement will continue to hold \emph{unless} the number of loops is a very very large number of similar magnitudes (and there is no way of disproving that something funny happens in this realm since we can't compute such a complicated Feynmann diagrams). Anyway, we now arrive at a prescription of identifying these constants. It goes as follows:

1. Start with two small numbers $m$ and $\epsilon_1$ as well as the large number $\Lambda$. While we \emph{know} that $\Lambda \gg m$ and $\Lambda \gg \sqrt{\epsilon_1}$, we do \emph{not} know the relation between $m$ and $\sqrt{\epsilon_1}$. 

2. Pick $\epsilon_2$ satisfying $\max (m^2, \epsilon_1) \ll \epsilon_2 \ll \Lambda^2$.

3. Pick $\xi$ satisfying $\xi \gg \Lambda^2/ \epsilon_2$. 

\subsection*{Can the values of the parameters be empirically assessed?}

So far we have found the relationship between "very small" ($m$, $\epsilon_1$ and $\epsilon_2$) and "very large" ($\Lambda$, $\xi$) parameters. But said relation can be satisfied if we make \emph{all} small parameters even smaller or \emph{all} large parameters even larger, as long as we change \emph{all} of them at the same time. So is there a way to determine what value do they "all" have? The way to do it is to find the value of "one" of them. Since $\epsilon_2$ is much larger than both $\epsilon_1$ and $m$, the latter two parameters are "hidden" due to "overarching" $\epsilon_2$. So the most logical way to do is to assess $\epsilon_2$. Now we remind the reader that the value of $\epsilon_2$ corresponds to the precision of Mensky's measurement. In particular, it determines a size of the "corridor" that path integral is confined to. This, however, is not entirely accurate: since the weight function is given by an \emph{integral},
\beq w = \exp \Bigg(- \frac{\epsilon}{2} \int d^4 x (\phi - \phi_{cl})^2 \Bigg)\eeq
it would be $e^{- \infty} = 0$ due to the infinite volume of spacetime, \emph{regardless} of how small $\epsilon$ might be. The way out of it is to say that the corridor "narrows down" asimptotically, while it becomes "wide" in some restricted region. Of course, the choice of the restricted region in which it becomes "wider" changes as we go through all possible paths: we first consider all possible paths that deviate from $\phi_{cl}$ throughout region $R_1$, then we consider all the paths that deviate from $\phi_{cl}$ throughout region $R_2$, and so forth. As discussed in more detail in \cite{Epsilon} the finite size of these regions corresponds to the finite lifetime of photons that we are talking about throughout this paper.  Now, in light of what we have just said, the weight function becomes
\beq w = \exp \bigg(- \frac{\epsilon}{2} V (\Delta \phi)^2 \bigg) \eeq
where $V$ is the volume of the region, and $\Delta \phi$ is the standard deviation of our \emph{assessed} value of \emph{average} $\phi$ throughout that region. Now, in order for a relevent contribution to path integral to be significant, we need $w > w_{min}$; or, in other words, 
\beq \exp \bigg(- \frac{\epsilon}{2} V (\Delta \phi)^2 \bigg) > w_{min} \eeq
This means that 
\beq \frac{\epsilon}{2} V (\Delta \phi)^2 < e^{- w_{min}} \eeq
In other words, making $V$ large would help us in measuring $\phi$ more precisely (in terms of $\Delta \phi$ being smaller) although this is at the expanse of the fact that we are mesuring the \emph{average} value of $\phi$ throughout the region rather than at the point. \emph{This is not to be confused with uncertainty principle:} uncertainty principle is independant of $\epsilon$ while what we do now depends on it. Now, we can use the knowledge of how well we can measure things to assess what $\epsilon_1$ and $\epsilon_2$ are, verify that $\epsilon_1 \ll \epsilon_2$ in fact holds (as in, we can measure $A^0$ a lot more precisely than $\vec{A}$ in a "preferred frame" of the model) and then use these results to make assessments about other parameters that are "tied together" to $\epsilon_1$ and $\epsilon_2$ per inequalities we derived in previous section. 

\subsection*{Propagators}

In the above section we have found that non-existence of longitudinal modes is only approximate. From exact point of view, they do exist, they just have extremely short lifetime. The fact that their lifetime is very small \emph{as opposed to} zero implies that we need "very small" modifications in the propagators as well. Indeed, the fact that $\epsilon_1 \neq \epsilon_2$,\emph{together with} the fact that $\epsilon$ has literal meaning in the propagator, is sufficient to tell us that we need propagator that includes both $\epsilon$-s, which the current propagator does not. First, we rewrite Eq \ref{ActionToInvert} as 
\beq S = \frac{A^{\mu} A^{\nu}}{2} (ag_{\mu \nu} + b k_{\mu} k_{\nu} + c v_{\mu} v_{\nu}) \eeq
where
\beq a = m_s^2-k^2 \; , \; b = 1 - \frac{1}{\xi} \; , \; c = m_t^2 - m_s^2 \label{abcDefinitions}\eeq
Now, the propagator takes the form 
\beq D^{\mu \nu} = a' g^{\mu \nu} + b' k^{\mu} k^{\nu} + c' v^{\mu} v^{\nu} + d' v^{\mu} k^{\nu} + e' k^{\mu} v^{\nu} \label{PropagatorGeneral} \eeq
where the coefficients $(a', b', c', d', e')$ are selected in such a way that 
\beq ( a' g^{\mu \rho} + b' k^{\mu} k^{\rho} + c' v^{\mu} v^{\rho} + d' v^{\mu} k^{\rho} + e' k^{\mu} v^{\rho})(ag_{\rho \nu} + b k_{\rho} k_{\nu} + c v_{\rho} v_{\nu}) = \delta^{\mu}_{\nu} \eeq 
If we evaluate the left hand side, while keping in mind $v^2=1$, this becomes
\beq \delta^{\mu}_{\mu} a'a + k^{\mu} k_{\mu} (a'b+b'a + b'b k^2 + e'b v \cdot k)  + v^{\mu}v_{\nu} (a'c+c'a+c'c + d'c k \cdot v) + \nonumber \eeq
\beq + v^{\mu}k_{\nu} (c'b v \cdot k + d'a + d'b k^2)  + k^{\mu} v_{\nu} (b'c k \cdot v + e'a + e'c) = \delta^{\mu}_{\nu} \eeq
Thus, we have five equations:
\beq a'a = 1 \eeq
\beq a'b + b'a + b'bk^2 + e'b v \cdot k = 0 \eeq
\beq a'c+c'a+c'c + d'c k \cdot v = 0 \eeq
\beq c'b v \cdot k + d'a + d'b k^2 = 0 \eeq
\beq b'c k \cdot v + e'a + e'c = 0 \eeq
Now, we already know the values of $a$, $b$ and $c$ from Eq \ref{abcDefinitions}; thus, we have five unknowns: $(a',b',c',d',e')$. This means that we can rewrite the above equation as 
\beq \left[ \begin{array}{ccccc}
 a & 0 & 0 & 0 & 0 \\
b & a +bk^2 & 0 & 0 & b v \cdot k \\
c & 0 & a + c & c k \cdot v & 0 \\
0 & 0 & b v \cdot k & a + bk^2 & 0 \\
0 & c k \cdot v & 0 & 0 & a+ c \end{array} \right]
  \left[ \begin{array}{ccccc}
a' \\
b' \\
c' \\
d' \\
e'  \end{array} \right] = 
\left[ \begin{array}{ccccc}
1 \\
0 \\
0 \\
0 \\
0 \end{array} \right] 
 \eeq
which has a solution
\beq a' = \frac{1}{a} \eeq
\beq b' = - \frac{b + \frac{bc}{a}}{(a+bk^2)(a+c)-bc(k \cdot v)^2} \eeq
\beq c' = - \frac{c+ \frac{bc}{a} k^2}{(a+cv^2)(a+bk^2) - bc(v \cdot k)^2} \eeq
\beq d' = \frac{ \frac{bc}{a} k \cdot v}{(a+c)(a+bk^2) - bc (v \cdot k)^2} \eeq
\beq e' = \frac{\frac{bc}{a} v \cdot k}{(a+bk^2)(a+c) -bc (k \cdot v)^2} \eeq
Substitutting this into Eq \ref{PropagatorGeneral}, we obtain
\beq D^{\mu \nu} = \frac{g^{\mu \nu}}{a} + \frac{- \Big(b+ \frac{bc}{a} \Big) k^{\mu} k^{\nu} - \Big(c+ \frac{cb}{a} k^2 \Big)v^{\mu}v^{\nu} + \frac{bc}{a} k \cdot v (v^{\mu} k^{\nu} + k^{\mu} v^{\nu})}{(a+c)(a+bk^2)-bc (v \cdot k)^2} \eeq
By substituting the values of $a$, $b$, and $c$ given in Eq \ref{abcDefinitions} and, furthermore, by substitutting Eq \ref{mepsilon} into $m_s$ and $m_t$, we obtain
\beq D^{\mu \nu} = \frac{g^{\mu \nu}}{m^2-k^2- i \epsilon_2} - \eeq
\beq - \frac{ \Big(1- \frac{1}{\xi} \Big) k^{\mu} k^{\nu} + i (\epsilon_1 + \epsilon_2) v^{\mu} v^{\nu} + \frac{i (\epsilon_1 + \epsilon_2)\Big(1- \frac{1}{\xi}\Big)}{m^2-k^2-i \epsilon_2} (k^{\mu} k^{\nu} + v^{\mu} v^{\nu} - k \cdot v (k^{\mu} v^{\nu} + k^{\nu} v^{\mu}))}{(m^2-k^2)\Big(m^2- \frac{k^2}{\xi} \Big) - i \epsilon_2 (m^2 -k^2) + i \epsilon_1 (m^2 - \frac{k^2}{\xi} \Big) + \epsilon_1 \epsilon_2 - i (\epsilon_1 + \epsilon_2) \Big(1 - \frac{1}{\xi} \Big) (v \cdot k)^2} \nonumber \eeq

\subsection*{Massive photons in GRW framework}

As was discussed in the Introduction, what we have been doing up till now was within a framework of one \emph{specific} quantum measurement model: namely, Mensky's restricted path integral. As a matter of fact, as has been shown in \cite{Epsilon}, the assertion that $\epsilon$ is finite \emph{implies} Mensky's path integral! So then the question is: what happens within the framework of other models? One logical example of alternative model to consider is GRW model: it has been shown in recent work, \cite{MenskyGRW}, that Mensky's path integral can emerge as a large-scale approximation of GRW model. Thus, it is logical to conclude that $i \epsilon$ is \emph{also} something that emerges only on large scales. So what happens on the scales where $\epsilon$ doesn't emerge? Are longitudinal photons more visible there?  

Let us first briefly review GRW model. According to that model (see \cite{GRW1}, \cite{GRW2}, \cite{GRW3}, \cite{GRW4}, \cite{GRW5}) wave function evolves unitarily, but that evolution is interrupted by "mini collapses" known as "hits". If "hit" number $k$ takes place at $t=t_k$, we can describe it  
\beq \psi (t_k^+, \vec{x}) = N (t_k^-, \vec{x}_k) e^{- \frac{\epsilon}{2} \vert \vec{x} - \vec{x}_k \vert^2} \psi (t_k^-, \vec{x}) \label{Hit} \eeq
where $N(\vec{x}_k; \psi (t_k^-))$ is a normalization constant needed to make sure that the \emph{new} wave function at $t^+$ is norm $1$, assuming that the \emph{old} one at $t^-$ was norm $1$. Furthermore, the point $\vec{x}_k$ is selected based on "weighted" probability given by 
\beq \sigma (\vec{x}_k = \vec{y}) = \frac{1}{N^2 (\vec{y}; \psi (t_k^-)} \label{HitProb}\eeq
 In light of the fact that $\epsilon$ is small, a single "hit" event, performed according to Eq \ref{Hit}, won't lead to significant change in $\psi$. However, a \emph{large number} of "hits" will amount to "taking a product" of several "wide" Gaussians that would produce a narrow Gaussian. It is easy to see that even if the centers of two or more Gaussians are different from each other, the product, which is also a Gaussian, will have well defined center. Thus, the "narrow Gaussian" we obtain after several "hits" will be centered around some well defined location, and the multiplication of wave function by such Gaussian will amount to the "collapse" of the latter.  Furthermore, one can show that if the selection of centers of wide Gaussian is done according to Eq \ref{HitProb}, then the location of the narrow Gaussian will obey Born's rule, thus making the "ultimate collapse" of wave function obey Born's rule as well. 

However, in light of the fact that wave function (due to Schrodinger's equation) diffuses between any two subsequent Gaussians, their effect might not accumulate into narrow Gaussian. But \emph{in certain settings} it does. For example, when an electron hits the screen, it interacts with all of the other electrons on the screen; this means that Gaussians acting on these \emph{other} electrons have indirect effect on the electron we are interested in. This magnifies the frequency of "relevant" Gaussians tremendously, leading to "narrow Gaussian" and, therefore, collapse. At the same time, from the pure formal point of view, the frequency of Gaussians applied to that electron is the same \emph{regardless} of whether it hit the screen or not: in both cases its very rare. Its the \emph{unitary evolution} of electron in question that is being altered due to Gaussians acting on \emph{other} electrons. Therefore, \emph{unlike} Copenhaggen interpretation, we do \emph{not} change "rules of the game" between measurement scenario and measurement-free one: in both cases we have Gaussians that come with the same frequency! In other words, any given electron is being measured \emph{at all times}, even if it is in the cosmos, and the strength of measurement is identical. The only thing that changes is the \emph{indirect} impact of measurement onto subsequent unitary evolution. 

This brings us back to the Mensky's path integral. While Mensky himself, at least some of the time, views his theory as emergent, I prefer to view it as fundamental. Thus, it is a \emph{fundamental property of nature} that it "continuously measures" the system with the "corridor" of a given width. This is emphasized by the connection between Mensky and $i \epsilon$ described in \cite{Epsilon} : since $i \epsilon$ is part of the fundamental setup that is idnependant of presence or absence of measuring device, the same should be true regarding Mensky's path integral. In light of this, it is no surprise that in \cite{MenskyGRW} the connection between the two models was found. In fact, these models say essentially the same thing: the only difference between them is that in Mensky's case the measurement is continuous while in GRW case it comes as a discrete set of events. Thus, it is no surprise that, on a time scale much larger than the time interval between those events, it will "look like" continuous measurement and $i \epsilon$ will emerge. 

Meanwhile, the path integral \emph{between} the events might still be well defined: in order to avoid the question about poles, all we have to do is to say that spacetime is discrete; this would imply that the contour of integration is discrete and, therefore, the "finite" set of points on the contour will "miss" the poll with absolute certainty. The problem that \emph{will} arize is that the discretized integral will approximate an average of two opposite directions of the contour, thus failing to explain why causality is future directed. Nevertheless, on a scale of \emph{several different} GRW hits the future time direction will emerge \emph{together with} $i \epsilon$. Physically, this is the result of the fact that each hit acts to the future and not to the past. We can then claim that we simply can't probe the time scale between two hits; the only scales we \emph{can} see are the ones in which time direction has already emerged.

Let us now go back to the question of massive photons. As we have shown in the previous chapters, the key reason why we don't see longitudinal modes is that $m \ll \epsilon_2$. Now, in GRW framework, both $\epsilon_1$ \emph{and} $\epsilon_2$ are \emph{emergent} parameters on a scale of \emph{several} hits; yet, $m$ is a fundamental parameter that is present in \emph{all} scales! Thus, on scale between two hits we have $m \neq 0$ yet $\epsilon_2 =0$, contradicting $m \ll \epsilon_2$. This means that we \emph{do} see longitudinal photons between two hits, but they are being "killed" \emph{by} the hits. 

Now, in order to understand why hits "kill" the longitudinal photons, we remind ourselves about our discussion in Introduction where we pointed out that "normal" values of $A_z$ and $A_t$ in the reference frame of the photon look "very large" in the reference frame of the observer; at the same time $A_x$ and $A_y$ are the same in both photon's and observer's frame, provided that photon moves in $z$-direction. Now, if the "hit" uses the observer's frame as the "preferred" one, then it would "measure" the strength of each component of $A$ in the observers frame up to the precision given by the width of the corridor. Now, "measuring" $A_t$ and $A_z$ \emph{in observer's frame} up to that precision amounts to "measuring" said parameters \emph{up to much higher precision} in photons frame (after all, they were "stretched" in observer's frame). This would essentially "kill" their physical impact. While the observer still thinks they are "large enough" due to "stretching", in the frame of photon itself they are tiny which is why they have no physical impact. 

What we have just said in the last paragraph violates letter and spirit of relativity. After all, relativity predicts that if A is stretched with respect to B, then B is \emph{also} stretched with respect to A. Saying that stretching of A contracts B (which is what we have said) is a non-relativistic argument. Now, the reason we have done it is that the entire GRW model is being imposed in non-relativistic way. After all, any given GRW hit is described as 
\beq \psi (A^{\mu}, t_k^+) = \psi (A^{\mu}, t_k^-) e^{- \frac{\epsilon}{2} d(A^{\mu}, A_k^{\mu})} \eeq
for some "distance function" $d$ (here, $\psi (A^{\mu})$ is the QFT replacement of $\psi (x)$ since now the variable we are "measuring" is $A^{\mu}$ rather than $x$). Now, if $d$ was defined in "covariant way", such as 
\beq d_{rel} (A^{\mu}, A_k^{\mu}) = (A^{\mu} - A_k^{\mu})(A_{\mu} - A_{k \mu}) \eeq
then $(A^0 - A_k^0)^2$ and $\vert \vec{A} - \vec{A}_k \vert$ would come with opposite signs. Thus, if \emph{one} of them is affected by Gaussian, the \emph{other} would be affected with \emph{inverse Gaussian} which would push it \emph{outward} rather than inward! This is something we don't want! Thus, \emph{similar to Mensky's case}, we break Laurentz covariance by putting \emph{the same} sign:
\beq d_{non-rel} (A^{\mu}, A_k^{\mu}) = \frac{\epsilon_1}{\epsilon_2} (A^0 - A_k^0)^2 + \vert \vec{A} - \vec{A}_k \vert^2 \eeq
Thus, it can only hold in a preferred frame. Said preferred frame is a frame of an observer \emph{as opposed to} a frame of a photon. Thus, if a photon \emph{happens to be} resting in the frame "preferred by" GRW hits, then it would have all three polarization modes. But if it moves with fast enough relative to such frame then each "hit" will, in non-relativistic way, "get rid of" the longitudinal modes that \emph{did} propagate \emph{right before} said hit took place. 

In light of this, the answer to the question of the lifetime of longitudinal photons is different. In particular, their life time has to either coincide, or exceed, the time interval between two subsequent hits. If each individual "hit" is very strong then the longitudinal photon lives until the next hit and then dies off immediately. On the other hand if the effect of each hit is small, then longitudinal photon can still live past each hit, but it will be weakened \emph{slightly}; then, it would \emph{eventually} die off on a scale of large number of hits, and in this case its lifetime can be approximated based on the equations we derived from $i \epsilon$ since said $i \epsilon$ \emph{would} emerge on large scales.

\subsection*{Massive photons in Bohmian case}

What we have said for GRW model can be restated in more general terms with regards to \emph{any} kind of measurement. We know from experience that small scales are quantum mechanical while large scales are classical. Yet, relativity tells us that longitudinal degrees of freedom can't possibly be "small" in all frames. At best, they are "small" in the frame of the photon, but they would become "large" in the frame of the observer that moves fast enough relative to the photon in question. Thus, in order to answer the question "is longitudinal photon classical or quantum mechanical" we have to have a "preferred reference frame" and say that the strenth of $A^z$ and $A^t$ \emph{relative to that frame alone} is what decides whether or not the photon is quantum mechanical. Now, a \emph{transverse} photon has a lot more options of "looking small" in the preferred frame than the longitudinal photon does. The transverse photon can move as fast as it wishes, as long as it makes sure to have $A^0=0$ \emph{in preferred frame}. On the other hand, a longitudinal photon is "forced" to be polarized in the directions involved in Lorentz boost and therefore its only "way" of looking small is to move slow enough; this of course is possible (after all it has a mass) but from pure statistical point of view the probability of moving with slow speed is small. Thus, the longitudinal photon will be statistically expected to violate the above condition which would make it "classical". Now since classical interactions, as such, do not exist but rather classical physics is emergent from quantum behavior, the lack of participation of longitudinal photon in quantum mechanical process essentially shuns its role from physics altogether. This phenomenon is famililar in other settings and is known as \emph{quantum Zeno effect}. Similarly to how electron in the atom is stopped from having \emph{any} physical significance (quantum \emph{or} classical) by being continuously measured, the same takes place with longitudinal photon. 

As one can see from the above argument, we have made no reference to any specific interpretation of quantum mechanics \emph{as long as} we assume that \emph{one of them} takes place. Therefore, we can now take something logically unrelated to the interpretations considered earlier -- namely Bohmian mechanics -- and see what would happen to massive photons in this case. 

Let us first review Bohmian model. While that model is unique for quantum mechanics case, there are several conflicting proposals on how to generalize it to quantum field theory (compare, for example, \cite{Bohm1}, \cite{Bohm2} and \cite{Bohm3}). Investigating each of these models is beyond the scope of this paper. We will, however, outline a general idea of what \emph{might} happen. The quantum mechanics version of Bohmian model (which is what ultimately inspired all the other models) is that the wave and the particle are two \emph{separate} entities with one way interaction: a wave acts upon a particle and not the other way around. The wave is like a river and the particle is like a piece of wood that floats in that river. If the river divides into two or more parts, then the water waves will flow on both sides and undergo interference if the two sides are to reunite. At the same time, the particle goes through either one side or the other, not both. It turns out that if the equation of motion of a particle is given by 
\beq \frac{d \vec{x}}{dt} = \frac{1}{m} \vec{\nabla} \; Im \ln \psi \eeq
then the probability of finding a particle can be shown to obey Born's rule. Now, the "measurement" is the result of wave function splitting into several non-overlapping branches. In this case, the particle will found in \emph{one} of them, while the other branch will no longer be relevent. This is called \emph{effective collapse}. The idea that they won't overlap again in the future might seem counter intuitive: after all, according to Schrodinger's equation the branches are supposed to "spread out" which would lead to an overlap; in fact, the purpose of \emph{either} Mensky \emph{or} GRW is to introduce some extra effects that would "narrow down" the wave function. On the other hand, Bohmians claim that none of these narrowing effects are needed and the branches would remain non-overlapping all on their own, per ordinary Schrodinger's equation! The justification of that claim is that the wave function in question lives in \emph{configuration space} rather than ordinary space. Thus, any given point in configuration space describes exact positions of every single molecule in the universe. Since it is "not likely" to change one set of "fingerprints" to another, one can aruge that it is equally unlikely to travel from one point in configuration space to another. In case of few particle quantum mechanics, the "point" in a 10-particle configuration is actually an $N-10$ dimensional hypersurface in the \emph{ultimate} configuration space. Thus, we \emph{do} have a "high probability" of traveling between any two such hypersurfaces, which is why the 10-dimensional wave function, in fact, spreads out. But this doesn't change the fact that the probability of traveling between two \emph{actual} points living in $N$ dimensions (as opposed to hypersurfaces in $N$ dimensions or points in 10 dimensions) is very small which is why the wave function in $N$ dimensional space does not spread out, which keeps the branches separate. 

Let us now apply it to massive photon. As we have pointed out, our ultimate goal is to say that the "measurement" restricts the fluctuations of \emph{all} space-like components of $A^{\mu}$ to the same scale (which is a function of $\epsilon_2$) in the "preferred frame" which would result in the loss of quantum nature of longitudinal components \emph{if} they are moving fast enough \emph{with respect to the above preferred frame}. Now, the above described "splitting into branches" happens as a natural part of unitary evolution \emph{as opposed to} any of the non-unitary processes "added" in Mensky's or GRW cases. Now, unitary process is derived from Lagrangian, which \emph{is} Lorentz invariant. So what is the source of violation of said invariance? One might first be tempted to point out that "preferred time direction" was needed in order to define a point in configuration space; thus, in QFT case a preferred foliation into spacelike hypersurfaces is needed in order to define a QFT state that evolves in time. This, however, is not completely satisfactory: while Bohmian model indeed requires the violation of $(\vec{x}, t)$ symmetry, one can still retain $(\vec{A}, A^0)$ one. While on the first glance it might appear that $x^{\mu}$ and $A^{\mu}$ are "tied together" through $F_{\mu \nu}$, the fact is that Bohmian model is designed in such a way that would make the prediction identical to the one given by Born's rule, and the latter would predict that the specific foliation into hypersurfaces needed to define things is irrelevent when it comes to ultimate result. 

A better way of answering the question where the violation of Lorentz covariance comes from is to say that while dynamics itself doesn't violate it, the initial conditions do. After all, from pure statistical point of view, we most of the physical objects that we see should move arbitrarily close to the speed of light; yet they don't! Now, the fact that air molecules, dust particles, and so forth, are moving slowly with respect to our own reference frame implies that longitudinal photons will have much larger impact on said dust particles than transverse photons would: after all, the dust particles will "think" that longitudinal photons have been "stretched". Now, the assertion made by Bohmians is that physical interaction leads to splitting of wave function into branches. This implies that there would be far more "splitting into branches" along "longitudinal" direction than there would be along "transverse" one. Therefore, within \emph{each} branch the longitudinal degrees of freedom of each photon will be almost exactly defined, while transverse ones would not. And, therefore, transverse degrees of freedom will be the only ones that will have a "room for oscillating" within each branch. 

What this tells us is that the longitudinal degrees of freedom would \emph{not} die. In fact, they would interact with dust particles \emph{even more} than transverse photons would! What would happen is that the interactions of longitudinal photons with various dust particles would be so frequent that their free propagation will "become irrelevent". Thus, due to our inability of tracking down the way they interact with dust particles, we perceive said longitudinal photons as simply random noize, while transverse ones are the only ones we can actually keep track of. 

Now, there is, in fact, a connection between Bohmian mechanics and what we said about Mensky. In particular, each "branch" of wave function can be confined inside the "corridor" that surrounds it. But, this corridor would be an \emph{emergent} consequence of the specific Schrodinger evolution (splitting into branches) rather than fundamental. Furthermore, the branches outside said corridor would continue to exist \emph{contrary} to what would have happened if they were destroyed by the non-unitary effects of $i \epsilon$. Nevertheless, the results will be similar to what one would expect \emph{if} $i \epsilon$ was present: the would-be effects of the "weight function" produced by $i \epsilon$ on the "relevent" branch "would have been" much smaller than its effects on ohter branches. Thus, in order to "disprove" the existence of $i \epsilon$ one would have to observe other branches. Since Bohmian particle can't see htem, it is "lead to believe" that $i \epsilon$ exist. Now, the non-unitarity effect of $i \epsilon$ is produced due to the fact that, while branches do split, they do'nt re-emerge. In actuality, since hte process is unitary, they \emph{would} re-emerge at a distant future, but not any time soon. On the other hand, the reason why they split at a \emph{recent} past, is due to the \emph{initial condition} that they didn't \emph{already} exist. Said initial condition results in time direction "away from" the spacelike hypersurface in which it was given, which happened to be in the past from right now. 

Thus, both violation of time reversal \emph{and} relativity are due to initial conditions in Bohmian case. This is to be contrasted with Mensky's and GRW cases where they were linked to the non-symmetric setup of the model. Likewise, in Bohmian case $\epsilon$ is to be determined from initial conditions as well: namely, we need to plug initial conditions into Schrodinger's equation in order to predict the width of a "typical" branch that the wave function will split into which, in turn, will be $\epsilon$. On the other hand, in Mensky and GRW cases, $\epsilon$ is a fundamental constant that is independant of initial conditions. 

\subsection*{Conclusion}

In this paper we have explained in \emph{continuous} fashion the \emph{false appearance} of discontinuity around photon mass $m=0$ (such as there being two polarization degrees of freedom when $m=0$ and three when $m \rightarrow 0$). It turns out that the conventional picture is not accurate because it assumes $\epsilon=0$. In this paper, when we set $\epsilon \neq 0$, we have found that the regime conventionally ascribed to $m =0$ actually corresponds to $m \ll \sqrt{\epsilon_2}$, while the one ascribed to $m \rightarrow 0$ corresponds to $m \gg \sqrt{\epsilon_2}$. Thus, even though $m$ is "small" it doesn't "approach zero" since it is bounded below by $\sqrt{\epsilon_2}$. The continuity is assured by the transition region around $m \approx \sqrt{\epsilon_2}$. In this paper we haven't studied this region because we view the absence of longitudinal modes as an empirical evidence that $m \ll \sqrt{\epsilon_2}$. Nevertheless, we have made it clear that \emph{if} such region did exist, we would have very short lived longitudinal photon. In fact, even in our case, longitudinal photons still exist; it is simply that their lifetime is \emph{much smaller} than the typical wavelength. In the $m \approx \sqrt{\epsilon}$ we would expect longitudinal photons to survive within a time period of single wavelength but not much more; and in $m \gg \sqrt{\epsilon}$ we would expect longitudinal photons to have a lifteime of several wavelength; yet their lifetime would still be finite. Furthermore we claim that the lifetime of \emph{transverse} photons is finite as well, due to the effect of $\epsilon$. The reason we have an illusion that trasverse photon has infinite lifetime is simply because other processes (such as its absorption into other particles) take place much faster than the annihilation the photon \emph{would still} go through even if it existed in the vacuum. 

The essential difference between transverse and longitudinal photon is simply that the former has much longer lifetime. The idea of making $\epsilon$ finite is part of our philosophy that \emph{all} physical constants that are typically assumed not to have definite value (such as gauge fixing term $\xi$, ultraviolet cutoff $\Lambda$ and so forth) in fact \emph{do} have well defined value, we simply don't have means of measuring it. In fact, it turns out that we had to refer to the concrete values of $\Lambda$ and $\xi$ (\emph{in addition to} the value of $\epsilon$) in order to compute lifetimes of the photons. 

As was shown in \cite{Epsilon}, there is an empirical way of estimating the value of $\epsilon$. Namely, it is a function of ratio of quantum and classical scales. This, of course, is more easily said than done: in order to measure $\epsilon$ one has to remove any other obstacles that stand in a way of making a measurement. But, \emph{if} one can measure $\epsilon$ this would be an upper bound on photon's mass -- the latter being \emph{much smaller} than the square root of the former. This would also give a new perspective on the experimentalists work in detecting photon mass. People other than myself continue to assume $\epsilon \rightarrow 0$ and, therefore, they are looking for the "typical" behavior of a massive particle (whose mass is much larger than $\sqrt{\epsilon}$. But in my case I assume that the photon mass, \emph{while still being finite} is much \emph{smaller} than $\epsilon$. Thus, the behavior of massive photon will be different from the behavior of other massive particles (in particular, its behavior will be dominated by attenuation rather than oscillation). This would imply a radically new set of experiments to try to search for the type of massive photon I am describing. Said massive photon will also require a preferred frame (as evident from the effects of $\epsilon_1$ and $\epsilon_2$ having the same sign and also from $\epsilon_1 \ll \epsilon_2$). Such "preferred frame" is hard to detect because it only has large effects on longitudinal photons that have very small lifetime. One can say that in spirit (although not in letter) the prediction is similar to cosmic microwave background which, likewise, has preferred frame.

Finally we have shown that $i \epsilon$, which is tied to a specific quantum measurement model (Mensky's) can be "mimicked" within the framework of other measurement models that don't explicitly include $\epsilon$. What happens in \emph{all} cases is that longitudinal photons get "stretched out" in "preferred frame" due to Lorentz transformation; as a result, they get "measured" a lot more "intensely" than the transverse ones do as their "projected" coordinate components continuously exceed the quantum-classical threshold. This constant measurement prevents them from playing physical role due to quantum Zeno effect. Now, the role of finiteness of $i \epsilon$ amounts to producing Mensky's version of quantum measurement. Therefore, the effect of $\epsilon_2 \gg m$ in "making $m$ unnoticeable" can be shown to be equivalent to "Mensky's version" of quantum Zeno effect that "stop" longitudinal photons from evolving.

\end{document}